# Molecular dynamics simulations of noble gas release from endohedral fullerene clusters

M.K. BALASUBRAMANYA*†, M.W. ROTH‡, P.D. TILTON† and B.A. SUCHY‡

† Department of Physical and Life Sciences, Texas A&M University-Corpus Christi, Corpus Christi, Texas 78412, USA.

‡ Department of Physics, University of Northern Iowa, Cedar Falls, Iowa 50614, USA.

We report the results of molecular dynamics simulations of the release of five species of noble gas atoms trapped inside a small cluster of fullerenes in the temperature range 4000K ≤ $T$ ≤ 5000K. We find that larger noble gas atoms are generally released at a slower rate and that helium is released considerably more rapidly than any of the other noble gases. The differing release rates are due not only to the differences in the size and mass of a given endohedral species but also because larger trapped atoms tend to stabilize the fullerene cage against thermal fluctuations. Unlike with the case of atoms entering fullerenes, we find that any atom escaping from the cage results in a window which does not close. Escape rate constants are reported and comparisons with experiment are discussed.

## 1. Introduction

Endohedral fullerenes, or carbon cages trapping atomic or molecular species, have received significant attention both experimentally [1-16] and theoretically [17-33]. Such systems with noble gas atoms trapped inside the molecular cage are formed while making fullerenes by passing an electric arc between carbon electrodes in an inert atmosphere of noble gases. Much interest has focused on the behavior of these systems for primarily two reasons. Endofullerenes are found terrestrially at meteor sites with [3]He trapped inside. Their study can throw light on their extraterrestrial origins, especially the prevalent conditions at the time of their formation [34]. Secondly, chemists have been interested in encapsulating noble gas atoms inside fullerene cages and study the interactions between the host and guest. Cross and Saunders have pioneered the insertion of [3]He into $C_{60}$ [35]. This endohedral molecule is chemically modified outside the cage in different ways and subjected to NMR analysis. Since every [3]He-labeled fullerene has a distinctive helium chemical shift, that shift can be used to pin down the structure of the derivative, as well as monitor the molecule's subsequent chemical transformations. [3]He NMR spectroscopy has thus become one of the most powerful tools for following fullerene chemistry. In addition to He, four other noble gases - Ne, Ar, Kr and Xe - have been inserted into fullerenes, making unusual and highly stable noble gas compounds in which no formal bond exists between the noble gas and the surrounding carbon atoms.

A very convenient way to experimentally probe an endohedral fullerene system is to raise its temperature until the encapsulated species is released, and to subsequently measure the concentration of the released species. Measurements have been made of the release of Ne from endohedral Ne@$C_{60}$ [15]. It is possible for the fullerene to release a Ne atom without the fullerene structure being destroyed, which is impossible if the Ne atom is simply pushed through the molecular cage, breaking the C-C bonds. Moreover, in the presence of impurities, the rate of release of trapped noble gas atoms is increased by orders of magnitude. A modified windowing mechanism has therefore been proposed, where the impurity (e.g. radical) adds to the cage and weakens fullerene bonds. The endohedral atom, according to this model, exits from the 'weak spot' of the cage, or its 'window', followed by the impurity detaching from the carbon atoms cluster, thus allowing reconstitution of the C-C bonds and the fullerene cage [15]. We conducted molecular dynamics (MD) simulations of the release of Ne from small Ne@$C_{60}$ clusters without impurities [33], and found that the simulations describe the system reasonably well as far as overall cluster dynamics and individual fullerene disintegration is concerned, but not when dealing with windowing at temperatures as low as seen experimentally. We strongly suspect that a modification of the character of the bonds in the MD simulations would be required to adequately describe the windowing suspected in real systems, but even then direct modeling of this process will require computational times of the order of the presently accepted age of the universe. Much remains to be understood regarding the process of release of endohedral species from fullerene systems.

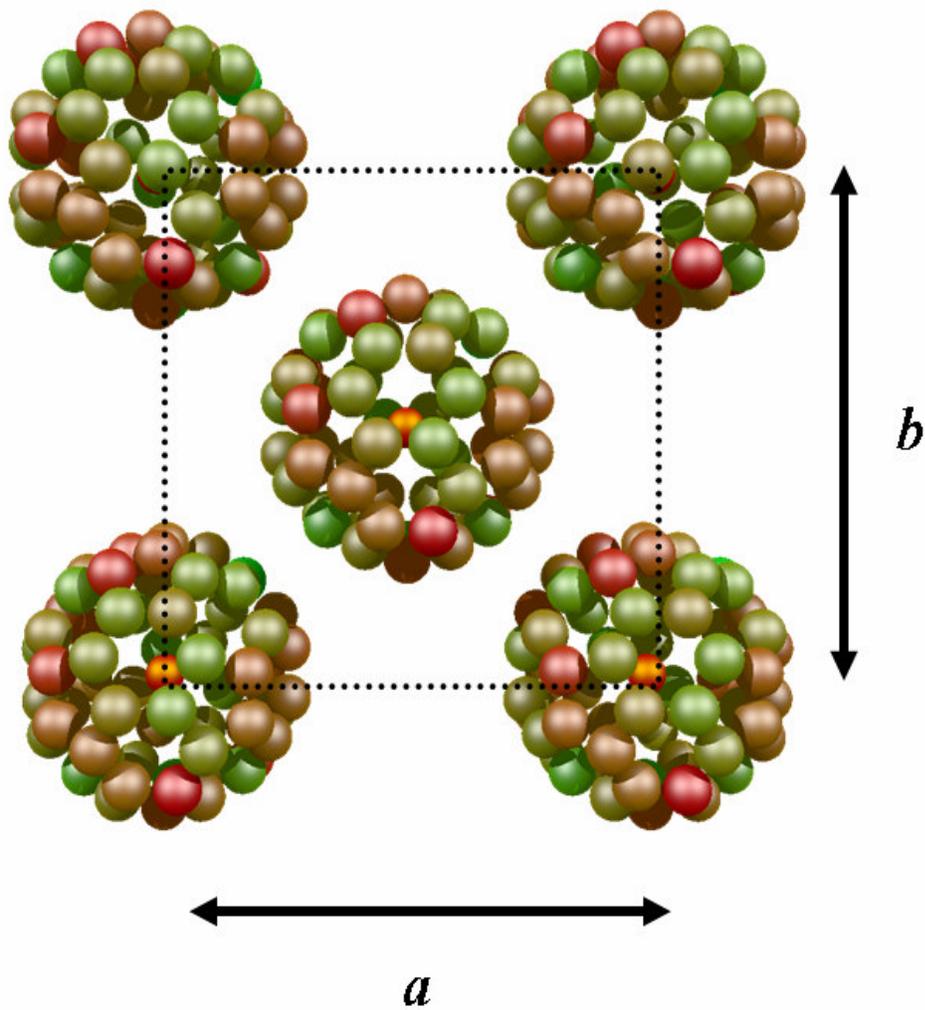

Figure 1. Initial conditions utilized for the simulations. The five fullerenes in the cluster form one face of an FCC fullerite unit cell with lattice constants $a = b = 14.4$ Å. The carbon atoms in each fullerene are colored green if they are closest to its center of mass, red if they are farthest away and a mixture of red and green if they are in between. The orange atoms inside the fullerene cages are encapsulated Ne atoms, and the relative atomic sizes, chosen for visual clarity, are not to scale.

Despite their limitations, MD simulations have provided a reasonable description of, and considerable insight into, fullerene systems [17-33]. Moreover, the noble gas atoms are a family of chemically similar species that differ mainly in their size and mass and, as such, they serve as ideal candidates for behaviour comparison in endofullerenes. The purpose of this study is to enhance our understanding of experimental and simulated endohedral release of noble gas atoms from fullerene systems by a comparative MD computer simulation. This study focuses on the release of five noble gas atoms encapsulated in $C_{60}$ clusters.

## 2. Computational Approach

The Ne@$C_{60}$ cluster chosen for this study has five endohedral fullerenes. The cluster size is chosen to be small because the process of release takes a substantial amount of simulated time. With a smaller cluster size it is possible to do many runs and obtain reasonable statistics. Moreover, in a real cluster containing many more fullerenes, as the temperature rises, smaller crystallites leave the cluster edge and it is likely that endohedral release happens in the gas phase from such small free crystallites. For this reason, periodic boundary conditions are not utilized; we wish to simulate small clusters where edge effects are important and cluster dissociation is not stifled. There is a very large reflecting box the cluster is kept in so that the system volume is constant. However none of the particles ever reflect off this wall in our simulation, so in actuality we implement free boundary conditions on the cluster. Above 257 K the fullerite crystal forms an FCC lattice. We model the initial configuration of the cluster at every temperature as one face of an FCC unit cell which has sheared off from the cell. As simulated time runs forward, the equations of motion are integrated using a standard Verlet algorithm with a time step $\Delta t=0.0005$ ps, and various structural averages, thermodynamic averages and relative atomic position distributions are calculated. In the temperature range 4000K $\leq T \leq$ 5000K, the results of 5 different runs are averaged at temperatures spaced 50 K apart, and temperature control is achieved by velocity rescaling for the carbon atoms and the noble gas population separately. Based on endohedral release times and the degree of equilibration of the system, each run is taken out to $2 \times 10^6$ time steps, or 1 ns. The initial configuration for the simulations is shown in Figure 1.

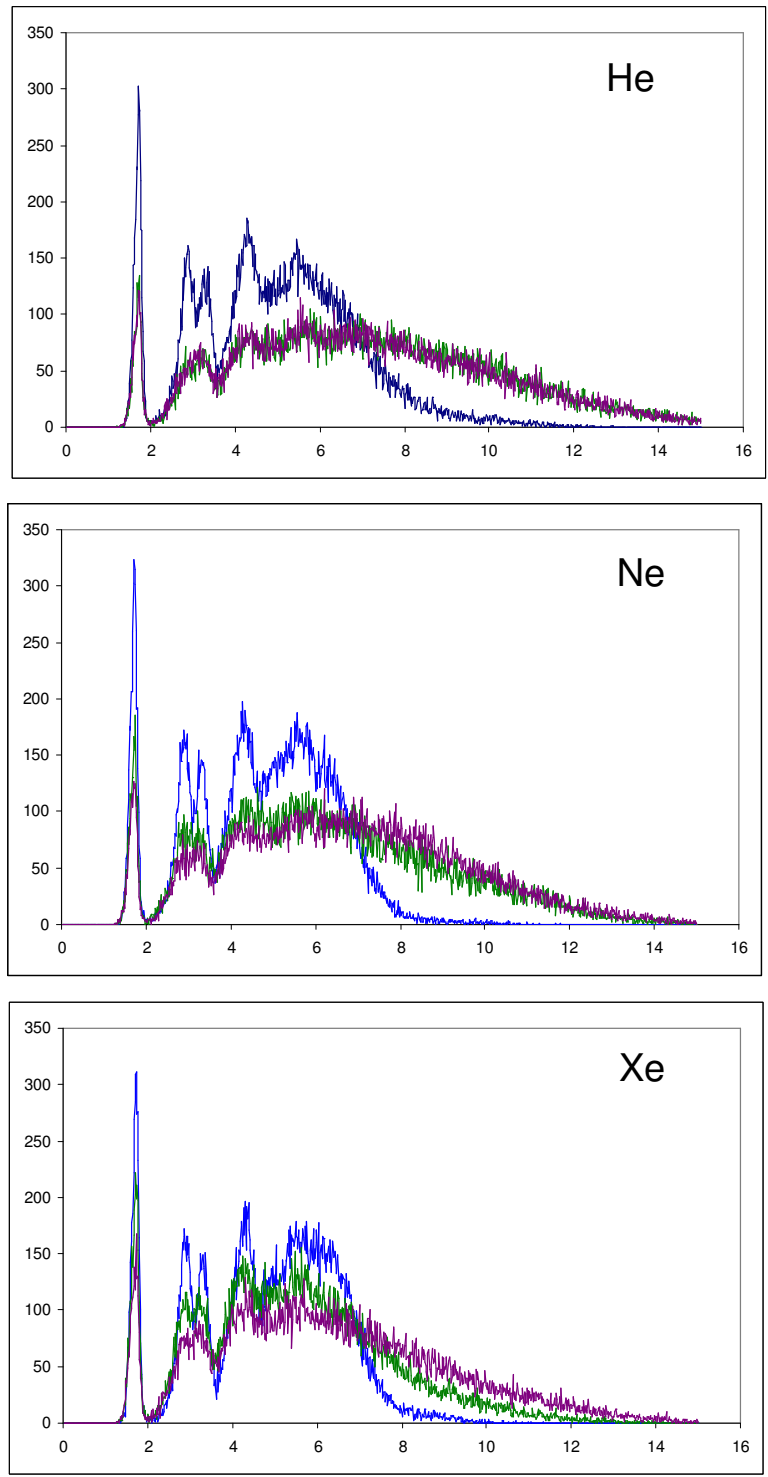

Figure 2. Fullerene pair distribution function $P_f(r_{ij})$ at $T = 4000$ K (blue), $T = 4500$ K (green) and $T = 5000$ K (purple) for He, Ne and Xe. The horizontal axes are in Angstrom and the vertical axes are arbitrary units; all axes are to the same scale.

There are several types of interaction potentials used in the simulations. The noble gas–noble gas potential as well as the noble gas-carbon potential are of a Lennard-Jones form,

$$u_{LJ}(r_{ij}) = 4\varepsilon_{ij}\left[\left(\frac{\sigma_{ij}}{r_{ij}}\right)^{12} - \left(\frac{\sigma_{ij}}{r_{ij}}\right)^{6}\right], \quad (1)$$

where the potential parameters for interaction between various species are given in table 1. Mixed interaction parameters are obtained with the use of Lorentz-Bertholot combining rules involving carbon-carbon parameters for the same potential as in equation 1. In addition, there is a non-bonded carbon-carbon interaction which is in a *modified* Lennard-Jones form [29],

$$\tilde{u}_{LJ}(r_{ij}) = \varepsilon_{CC}\left[\left(\frac{\sigma_{CC}}{r_{ij}}\right)^{12} - 2\left(\frac{\sigma_{CC}}{r_{ij}}\right)^{6}\right] \quad (2)$$

whose parameters are also shown in table 1.

Table 1. Parameters for the non-bonded Lennard-Jones (LJ) interaction potentials. The interactions with asterisks (*) are not used explicitly in the simulations because they are for a standard LJ interaction, not the modified one actually used in this study. They are used only in the combining rule relationships to get noble gas-carbon interaction parameters in the LJ potential.

| Species | $\varepsilon_{ij}$(K) | $\sigma_{ij}$ (Å) |
|---|---|---|
| He-He | 10.80 | 2.57 |
| Ne-Ne | 36.68 | 2.79 |
| Ar-Ar | 120.0 | 3.38 |
| Kr-Kr | 171.0 | 3.60 |
| Xe-Xe | 221.0 | 4.10 |
| C-C* | 28.00* | 3.40* |
| C-C | 34.839 | 3.805 |

The non-bonded carbon–carbon potential parameters given in table 1, and used in equation 2, are not derivable from the potential and the parameters in equation 1. The asterisked parameters (for the traditional Lennard–Jones interaction) apply to fullerene adsorption onto graphite [36] while the parameters for the modified Lennard–Jones potential for atomic carbon–carbon interactions apply to non-bonded fullerene carbons [29]. We have used the modified potential developed by Guo et al [38] from first principles for sp$^2$ carbon centers by fitting experimental lattice parameters, elastic constants and phonon frequencies for graphite. They have used this potential to successfully interpret and predict various properties for fullerene molecules and crystals, including data from vibrational spectroscopy, NMR, STM and crystal structure analysis.

The carbon-carbon bonded interactions are modeled by Brenner's empirical extended bond-order potential [37],

$$V_R(r_{ij}) = f(r_{ij})\frac{D_e}{S-1}\exp\left[-\beta\sqrt{2S}(r-R_e)\right]$$

$$V_A(r_{ij}) = f(r_{ij})\frac{D_e S}{S-1}\exp\left[-\beta\sqrt{\frac{2}{S}}(r-R_e)\right]$$

$$f(r_{ij}) = \begin{cases} 1, & r < R_1 \\ \frac{1}{2}\left[1+\cos\left(\frac{(r_{ij}-R_1)\pi}{(R_2-R_1)}\right)\right], & R_1 < r_{ij} < R_2 \\ 0, & r_{ij} > R_2 \end{cases} \quad\quad (3a)$$

which has parameters that are fit to various energetics of hydrocarbons, diamond and graphite. In equations (3a), $V_R$ and $V_A$ are the repulsive and attractive potential energy terms, respectively, which are essentially modified Morse potentials. The screening function $f(r_{ij})$ restricts the interaction to nearest neighbors as defined by the values for $R_1$ and $R_2$. In addition, the Brenner potential takes bonding topology into account with the empirical bond order function $\bar{B}_{ij}$ given by the relationships

$$B_{ij} = 1 + \left[\sum_{k\neq i,j}^N G_C(\theta_{ijk})f(r_{ik})\right]^{-\delta}$$

$$G_C(\theta) = a_0\left[1+\frac{c_0^2}{d_0^2}-\frac{c_0^2}{d_0^2+(1+\cos\theta)^2}\right]. \quad (3b)$$

$$\bar{B}_{ij} = \frac{B_{ij}+B_{ji}}{2}$$

Here the three-body bond angle is defined as

$$\theta_{ijk} = \cos^{-1}\left(\frac{\vec{r}_{ji}\cdot\vec{r}_{jk}}{r_{ji}r_{jk}}\right), \quad (3c)$$

where $\vec{r}_{ab}$ is the displacement vector from carbon atom $a$ to carbon atom $b$. Variations of the Brenner potential have been used for many different types of carbon allotrope simulations, as the empirical bond order function controls clustering to some extent. For example, Yamaguchi et al.[24-28] do not include information from the conjugate compensation term [37] because with it the potential would not adequately apply to small clusters having non–terminated carbons. Since we are dealing initially with complete fullerenes which eventually break up, we disregard the compensation term as well. The total carbon-carbon interaction is a sum over all bonded and non-bonded interactions:

$$u_{CC} = \sum_{i=1}^{N}\sum_{j>i}^{N}\left\{\left[V_R(r_{ij}) - \bar{B}_{ij}V_A(r_{ij})\right] + P_{ij}u_{LJ}(r_{ij})\right\}. \quad (4)$$

Here $P_{ij}$ is a screening function [34] which we implement by creating bonded and non-bonded neighbor lists. All carbon-carbon bonded potential parameters are given in table 2.

Table 2. Parameters for the bonded carbon-carbon Brenner interaction potential.

| Parameter | Value |
|---|---|
| $D_e$ | 73333.33 K |
| $\beta$ | 1.5Å$^{-1}$ |
| $S$ | 1.29 |
| $R_e$ | 1.315Å |
| $R_1$ | 1.750Å |
| $R_2$ | 2.000Å |
| $\delta$ | 0.80469 |
| $a_0$ | 0.011304 |
| $c_0$ | 19 |
| $d_0$ | 2.5 |

**3. Results and Discussion**

As the simulations proceed, and as endohedral atoms are released, the results of five simulations are averaged at each temperature and the number of atoms still contained within the cluster is fit to a function

$$N(t) = N_0 e^{-kt} \quad (7).$$

Representative results for all species at two selected temperatures, at the extremes of the temperature range chosen for the study, are shown in table 3. It seems clear that He is a special case, releasing much more rapidly than any of the other noble gases. Moreover, it is apparent in table 3 that the release behavior of Kr and especially Xe are affected much less by increasing temperature than is the case with the other noble gas species. Such behavior is further confirmed in table 4, where the escape constants themselves, for each species, are fit to exponential models of the form

$$k(T) = Ae^{KT} \qquad (8).$$

Table 3. Average release rate constants $k$ for all noble gas species at two representative temperatures at the extremes of the temperature range studied. Uncertainties are on the order of 40% of the mean values obtained by simulations.

| Species | $k\ (T = 4100K),\ \text{ps}^{-1}$ | $k\ (T = 4900K),\ \text{ps}^{-1}$ |
|---|---|---|
| He | .00359 | .0784 |
| Ne | .00029 | .0086 |
| Ar | .00024 | .0088 |
| Kr | .00027 | .0035 |
| Xe | .00019 | .0016 |

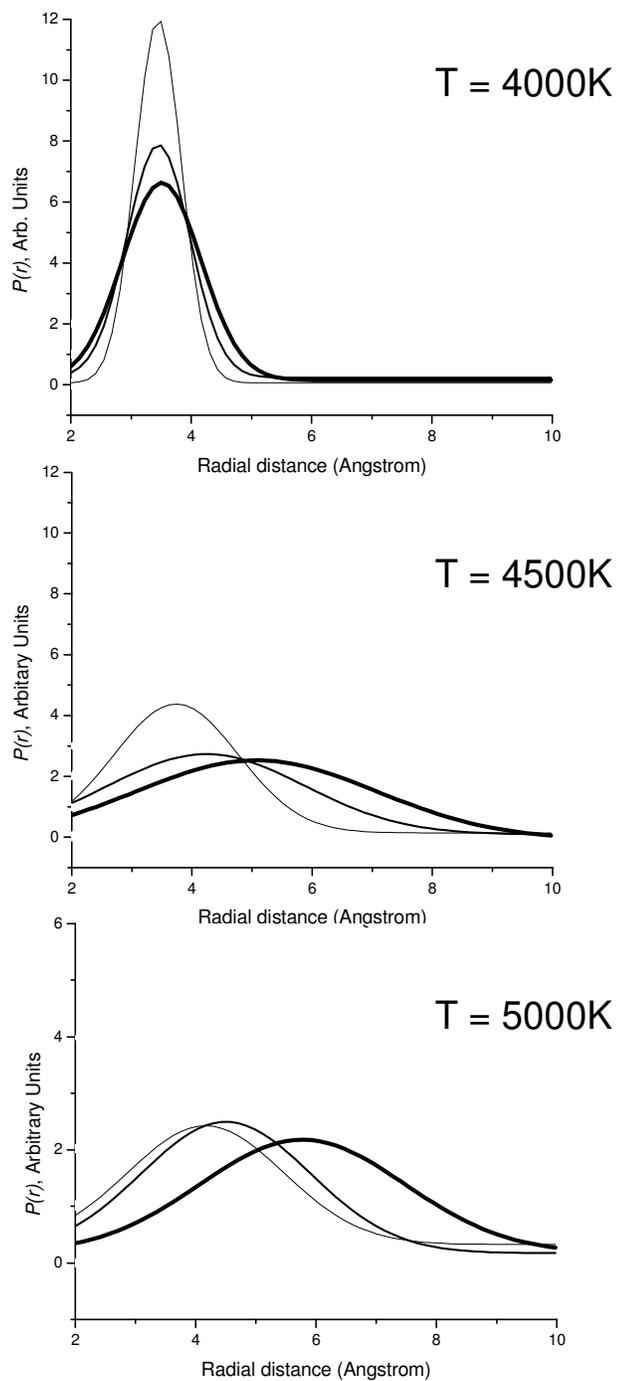

Figure 3. Radial fullerene probability distribution $P(r)$ for He (darkest), Ne (medium) and Xenon (lightest) at $T$ = 4000K, 4500K and 5000K. Horizontal axes are to the same scale but the vertical axes are chosen so as to maximize clarity.

Table 4. Constants $K$ for best fits to average release rate constants $k$ as functions of temperature $T$ according to the model $k = A\, e^{KT}$. Uncertainties are on the order of 30%.

| Species | $K$ (K$^{-1}$) |
|---------|----------------|
| He | 0.0042 |
| Ne | 0.0044 |
| Ar | 0.0047 |
| Kr | 0.0032 |
| Xe | 0.0025 |

It is clear that the values for $K$ are similar for the smallest three noble gases and then decrease somewhat for Kr and considerably for Xe. Given the uncertainties involved as well as other data generated by our simulations, the behaviour difference exhibited by the two heavier noble gases seems credible.

To better understand the release behavior of the various endohedral species examined it is necessary to look at the dynamics of the system in detail. The general trend of larger endohedral species being released more slowly can be understood in terms of three considerations: i) larger species have more difficulty exiting similar sized windows and also have more difficulty in making exit windows; ii) heavier species move more slowly than lighter ones at a given temperature, and iii) the larger and heavier species stabilize the fullerene cage against thermal fluctuations and hence defect formation. The first two considerations are self–evident but the last one is borne out by the simulation results. Figure 2 shows the fullerene pair distribution function $P_f(r_{ij})$ for He, Ne and Xe at temperatures $T$ = 4000K, 4500K and 5000K. $P_f(r_{ij})$ is the calculated frequency of occurrence of separation between two specific carbon atoms in different fullerenes being between $r_{ij}$ and $r_{ij} + \Delta r_{ij}$, divided by the normalizing factor $2\pi r_{ij}\Delta r_{ij}$. At low temperatures the fullerene cage shows solid–like ordering for all species. Likewise, at high temperature the cages for all species show considerable disorder. However, there are more differences between the $T$ = 4500K and $T$ = 5000K curves for Xe than for any other species, and the difference narrows for Ne and disappears for He. That is, at intermediate temperatures for endohedral release, the cages containing heavier atoms are considerably more organized than for lighter ones. In fact, for He the character of the cage at intermediate temperatures is indistinguishable from that at high temperatures.

In addition to examining atom–atom separations it is important to consider the radial localization of the cage. Figure 3 shows best Gaussian fits to $P(r)$ for the same systems shown in figure 2. The curves shown in figure 3 are the calculated probability that a carbon atom is at a distance $r$ from the center of its fullerene molecule. It is evident that, at all temperatures, cages containing He are much more disordered than for any of the

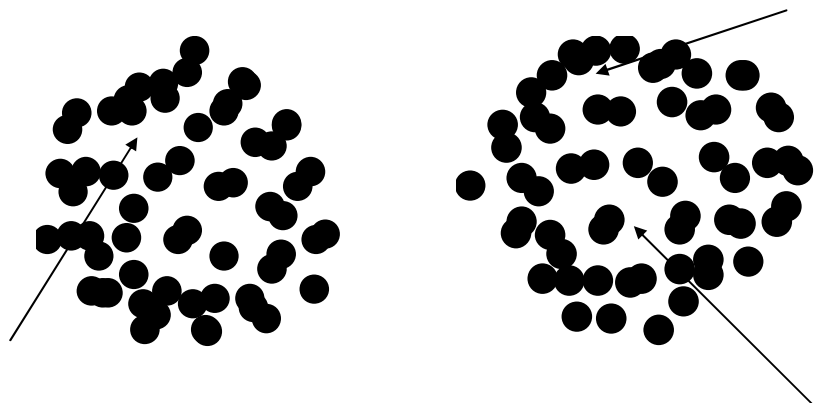

At $T = 4000K$ Note small, dynamic windows; the noble gas is not Released.

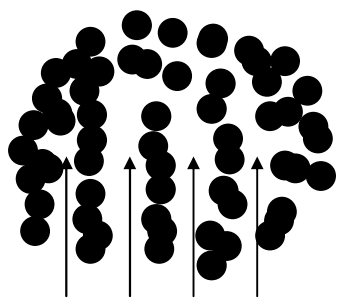

At around $T = 4200K$ note the permanent windows That become more prevalent as temperature increases. The noble gas has not escaped yet.

This part unfolded to make a large, permanent window, Releasing the noble gas.

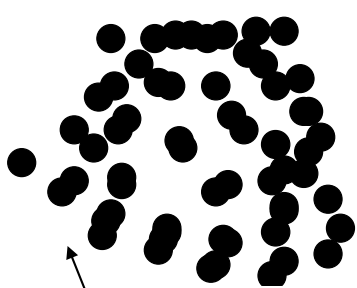 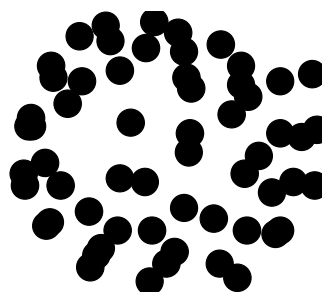

Ca. $T = 4500K$

This part unfolded and opened to make a large, permanent window. The noble gas was released.

Figure 4. Individual fullerene snapshots at various stages of the release process, which happen at slightly different temperatures for different species. The upper two pictures at $T = 4000K$ show small windows which dynamically open and close; the middle picture for $T = 4200K$ shows permanent windows which do not result in escape in our simulations and the lower two pictures shows the permanent damage to the cage resulting from endohedral exit. The results shown here are for Ne but the release stages are similar and representative of other encapsulated noble gases.

other species. At low and intermediate temperatures, cages containing larger and heavier species are more localized than for lighter species. At high temperatures, the difference between Kr and Xe cases has become statistically nonexistent, because larger atoms create larger (permanent) exit windows (as will be discussed later), imparting more destruction to the cage. The distributions in figures 2 and 3 taken together show that not only is there more solid–like order present in the fullerene cage for larger and heavier noble gas species, but also that at low and intermediate release temperatures these larger species force specific radial localization of the cage. It then seems reasonable to conclude that the stronger coupling to the cage that the larger and heavier noble gas atoms exhibit stabilize and localize it, which retards the trapped atom's release. It has also been observed in earlier simulations [30] that larger encapsulated species stabilized the fullerene cage at temperatures well below those at which release would take place.

One of the premier features of computer simulations (especially MD) is the ability to examine microscopic dynamics and processes stage by stage, and we are able to inspect the windowing and release details in the systems under study. Figure 4 shows representative snapshots of the fullerene cage at various points in the release process. Even though some release takes place at $T = 4000K$ we find that when the endohedral atoms are not released, small windows open and then reform, as shown in the top two snapshots. As temperature increases, the windows become larger and do not close, as shown in the middle snapshot of Figure 4. When escape takes place, the guest atom opens up, or sometimes unfolds, the cage as shown in the bottom two snapshots. In our simulations, endohedral release is always accompanied by such partial destruction of the cage, which does not heal. This helps explain why the release behavior of Kr and especially Xe are affected less by increasing temperature than for the other species: their exit maneuvers depend less on the thermally induced cage defects already present, and more on the guest atom's ability to make a defect large enough to exit through.

The difference in character between the dynamic windowing at lower temperatures and permanent windowing at higher temperatures is clearly seen in Figure 5. It shows details of the first neighbor peaks of Figure 2, and also includes data for all temperatures between 4000K and 5000K examined. The first neighbor peak shows the frequency of bonded pairs in the fullerene cage. Therefore, windowing will result in a lowering of this first peak, and more frequent windowing or larger windows will increase that signature. It is clear that for Xe and Ne the curves are bunched into two groups. Inspection of our data reveals that the group of curves with the higher peaks corresponds to systems with more dynamic (small) windows than larger permanent ones and the group with lower peaks has a large frequency of permanent windows. Figure 5 also shows that the cages containing He show much more destructive windowing at earlier temperatures, supporting the conclusions mentioned earlier that the cages containing He are much less stabilized than those containing other noble gases.

As observed in our previous study involving pure Ne@$C_{60}$ clusters, the results here differ significantly from those obtained (1) experimentally and (2) computationally at lower temperatures [15]. We attribute this to two reasons. Firstly, our simulation included no exhohedral noble gas atoms or other impurities whose collisions with the fullerene cage and attachment would have contributed to windowing. Thermal agitation is the sole source of windowing in this simulation. A temperature dependent description of the bond integrity as affected by any exohedral species will be required for a more realistic description, but no such model currently exists. Secondly, the structural and dynamic character of the fullerene cage and the onset of defect formation as revealed by our simulations are similar to those obtained from other simulations [19,20], but the onset of cage disintegration in this study is seen at different temperatures – higher in some cases and lower in others. It is also interesting to note, however that some theoretical calculations of fullerenes [31] places the cage disintegrating temperature at between $T = 4000K$ and $5000K$, which agrees with our work but differs considerably from what is obtained by various MD simulations [20, 24-28]. Thirdly, to simulate guest atom escape at lower temperatures the escape process will have to be artificially accelerated without compromising the integrity of the physics of the model, because in order for the endohedral population to become half of its initial value, the simulation will have to run for one half life which, even taking an underestimate of $t_{1/2} = 10$ hours, would require $7.2 \times 10^{19}$ steps using a time step of 0.0005 ps, translating to a real time commitment (in the case of this study) of the order of the age of the universe. Our simulations are able to reproduce many of the static properties of fullerite clusters [29] such as its structure, binding energy and dissociation temperature. As far as fullerene melting and disintegration, the current state of affairs suggests that MD simulations can give reasonable insight into the dynamics exhibited by such systems, especially when comparing the system's behavior when different atomic species are included in the simulations and all that is changed is the potential parameters. However, it is widely known that the potentials used for MD simulations do not numerically agree with experimental results and will not be able to accurately reproduce the melting and disintegration temperatures until a potential better describing the bond integrity is formulated. Considering all these factors, and that our results for He show that the cage can be considerably affected by an endohedral species, comparison of our work to that of other simulations reveals reasonable agreement. Comparison with experiment leads us to suspect that there are significant challenges in realistically simulating the type of noble gas release experimentally observed at temperatures below 1000K, where release half lives are on the order of hours.

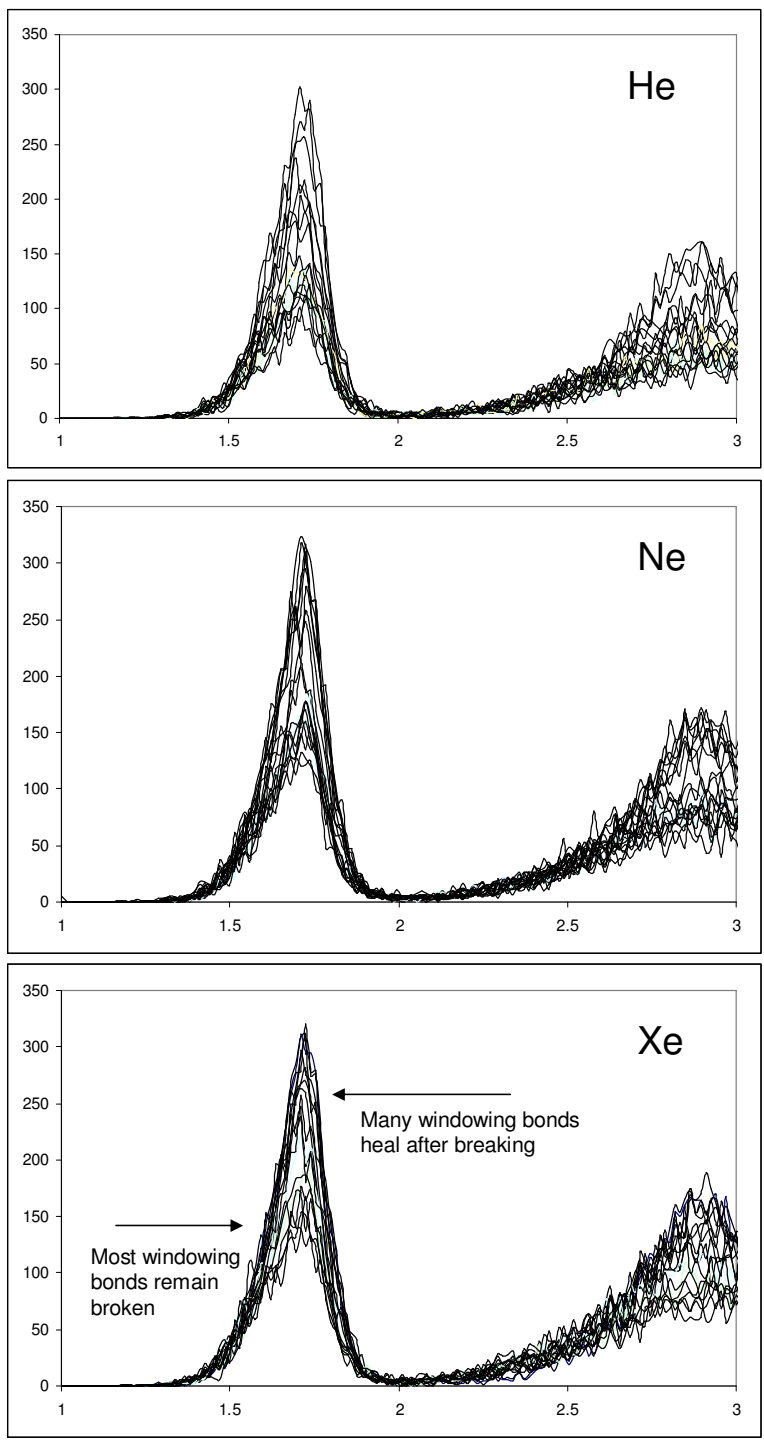

Figure 5. Fullerene pair distribution function $P_f(r_{ij})$ from $T = 4000$ K to $T = 5000$K in steps of 50K for He, Ne and Xe. Detail is on the first neighbor peak; all axes are to the same scale and have units identical to those in Figure 2.

**Acknowledgements**

M.W.R acknowledges useful discussions with J. Che and R.J. Cross, as well as a University of Northern Iowa summer fellowship in 2004. M.K.B. acknowledges support for molecular dynamics simulations by the U.S. National Science Foundation's Major Research Instrumentation program, under the Division of Computer and Network Systems' grant numbered 0321218.